\documentclass[12pt, a4paper] {article}

\usepackage{amsfonts}
\usepackage{verbatim}
\usepackage{color}
\usepackage{amsmath}
\usepackage{graphicx}
\usepackage{bm}
\usepackage{amssymb}
\usepackage{latexsym}
\usepackage{mathrsfs}
\usepackage{textcomp}
\usepackage{ulem}
\newcommand{\Lim}[1]{\raisebox{0.5ex}{\scalebox{0.8}{$\displaystyle \lim_{#1}\;$}}}

\begin{document}

\begin{center}
{\bf\large{Supervariable approach to particle on a torus knot: A model for Hodge theory} }

\vskip 1.5 cm

{\sf{ \bf Anjali S and Saurabh Gupta}}\\
\vskip .1cm
{\it Department of Physics, National Institute of Technology Calicut,\\ Kozhikode - 673 601, Kerala, India}\\
\vskip .15cm
{E-mails: {\tt anjalisujatha28@gmail.com, saurabh@nitc.ac.in}}
\end{center}
\vskip 1cm

\noindent
{\bf Abstract:}  We analyze a particle constrained to move on a $(p,q)$-torus knot within the framework of supervariable approach and deduce the BRST as well as anti-BRST symmetries. We also capture the nilpotency and absolute anti-commutativity of (anti-)BRST symmetries in this framework. Further, we show the existence of  some novel symmetries in the system such as  (anti-)co-BRST, bosonic, and ghost scale symmetries. We demonstrate that the conserved charges (corresponding to these symmetries) adhere to an algebra which is analogous to that of de Rham cohomological operators of differential geometry. As the charges (and the symmetries) find a physical realization with the differential geometrical operators, at the algebraic level, the present model presents a prototype for Hodge theory.

\vskip 1.5 cm



\newpage
\section{Introduction}
Gauge symmetries - a cornerstone in the reign of theoretical physics,  play a crucial role in underpinning our understanding of fundamental interactions of nature and have far-reaching upshots in the mathematical formulation of many significant physical theories.  One of the salient characteristics of any gauge theory is that they are endowed with unphysical degrees of freedom \cite{gauge_Henn}. Thus, appropriate gauge-fixing conditions have to be designated to remove these redundant degrees of freedom; however, this procedure destroys the manifest gauge invariance \cite{rotor}. Thereupon, Becchi-Rouet-Stora-Tyutin (BRST) formalism provides a powerful tool for the covariant quantization of a gauge theory \cite{brst1, brst2, brst3}. The two key features that delineate the BRST symmetry transformations (and corresponding charges) are: (i)- nilpotency and (ii)- absolute anti-commutativity.  Moreover, the geometrical origin and interpretation of these aforementioned characteristics are elucidated within the superfield framework \cite{super1, super2, super3, sup3, sup4, sup5}. Pursuant to this formalism, any $D$-dimensional gauge theory is generalized onto a $(D, 2)$-dimensional supermanifold parametrized by a pair of Grassmannian variables to derive (anti-)BRST symmetries wherein the horizontality condition offers a decisive role. Besides, in the superfield approach, the translational operators along the Grassmannian directions render a geometrical basis for the symmetries and the corresponding charges.

The present endeavor deals with a system of particle constrained to move on a torus knot.  The notion of knot theory generated great intrigue across widely disparate disciplines \cite{knot}. The proposition of knot theory first appeared in Lord Kelvin's hypothesis where atoms are characterized as knotted vortex tubes in ether, which in turn, motivated its emergence in diverse physical, chemical and molecular biological systems \cite{knot, wit}. To name a few, knot-like structures appear as stable finite-energy solutions in local $3$-dimensional Lagrangian of field-theoretic models \cite{fadd} and in the isofrequnecy contours generated by the topological surface states in $2$-dimensional Chern insulator \cite{chern_insu}. These knot-like structures also play a beneficial role in ascertaining certain gauge invariant observables in quantum gravity approaches \cite{knt_rovelli}. Moreover, the classical and quantum mechanical aspects of a particle confined on a torus knot have been explored to procure the energy spectrum \cite{int2}.  On top of that Hamiltonian analysis has been carried out for the same system to offer a description on dynamics and symmetries \cite{ghosh}.

In our recent work, we have constructed the gauge invariant theory of particle constrained to move on a $(p,q)$-torus knot via symplectic gauge invariant formalism \cite{gupt}. Here, we have imposed a constraint condition $(p\theta + q\phi) \approx 0$, where $p$ and $q$ are mutually prime numbers, into the Lagrangian with the aid of the Lagrange multiplier (say $\lambda$). Subsequently, by employing the symplectic gauge invariant formalism \cite{neto, nejad}, we have derived the first-order gauge invariant Lagrangian as \cite{gupt}
\begin{eqnarray}
L_f = \dot{\eta}P_\eta + \dot{\theta}P_\theta + \dot{\phi}P_\phi - \frac{(\cosh\eta - \cos\theta)^2}{2ma^2}\left( P_\eta^2 + \frac{\left( qP_{\theta}- pP_{\phi}\right)^{2}}{\left( p^{2}\sinh^{2}\eta + q^{2}\right)} \right) + \lambda(p\theta + q\phi) ,
\end{eqnarray}
where $\eta$, $\theta$, $\phi$ represent the generalized coordinates and $P_{\eta}$, $P_{\theta}$, $P_{\phi}$ denote the corresponding canonical momenta. The constraint analysis, {\it \`{a} la} Dirac, establishes the presence of two first-class constraints $ \chi_{1} \equiv P_{\lambda} \approx 0$ and $\chi_{2} \equiv (p\theta + q\phi) \approx 0$. These first-class constraints generate following infinitesimal gauge symmetry transformations \cite{gupt}
\begin{eqnarray}
\delta P_{\theta} = p\kappa, \quad \delta P_{\phi} = q\kappa, \quad \delta \lambda = \dot{\kappa}, \quad \delta [\eta, P_{\eta}, \theta, \phi] =0,
\end{eqnarray}
where $\kappa$ is the time-dependent infinitesimal gauge parameter.  Moreover, this newly constructed gauge invariant theory can be rephrased as a quantum system to possess a more generalized BRST symmetry (cf. section 2 for more details).

The primary motive of our present investigation is to derive the (anti-)BRST symmetries within the framework of supervariable approach. The second motive is to perceive {\it{new}} symmetries in the model and further provide a prototype for the Hodge theory. This may be realized as a bridge connecting the symmetries in the Lagrangian framework to the mathematical elements of differential geometry \cite{hod4, hod1, hod2}.  Third, we wish to provide a geometrical interpretation for the salient facets of (anti-)BRST symmetries, namely nilpotency and absolute anti-commutativity, by utilizing the supervariable approach.

The contents of the paper are organized as follows: Section 2 provides a brief account of the (anti-)BRST symmetries and charges for a particle confined to move a torus knot. We deduce the off-shell nilpotent and absolutely anti-commuting (anti-)BRST symmetries within the framework of supervariable approach in section 3. Further, section 4 captures the key properties of (anti-)BRST symmetries in the context of supervariable approach. Our section 5 comprises the analysis of (anti-)co-BRST symmetries and corresponding charges.  We then derive the bosonic and ghost scale symmetries along with their conserved charges in section 6. In section 7, we demonstrate that our conserved charges (and corresponding symmetries) provide a physical realization of the cohomological operators of differential geometry at the algebraic level. Finally, we summarize our results in section 8.

\section{Preliminaries}
We consider the reformulated gauge theory of a particle constrained to move on a $(p,q)$-torus knot within the BRST formalism. The BRST invariant first-order Lagrangian $(L_b)$ is constructed by including a BRST invariant function into the gauge invariant Lagrangian ($L_{f}$) as described below (cf. \cite{gupt}) as 
\begin{eqnarray} \label{L_brst}
 L_b &=&  L_f +  s_{b}\left[ i\bar{c}\left( \dot{\lambda} +  P_\theta + P_\phi  - \frac{1}{2}b \right)\right].
\end{eqnarray}
The BRST symmetries ($s_{b}$),  where gauge parameter is being replaced by (anti-)ghost variables, are listed below
\begin{eqnarray} \label{brst}
	&& s_{b}\lambda \;=\; \dot{c}, \quad 	s_{b} P_{\theta} \;=\; cp, \quad s_{b} P_{\phi} \;=\; cq, \quad	s_{b}\bar{c} \;=\; ib, 	   \\ \nonumber
	&& s_{b} c \;=\; 0, \quad s_{b} b \;=\; s_{b} \eta \;=\;  s_{b} P_{\eta} \;=\;  s_{b} \theta \;=\;   s_{b} \phi \;=\;  0.
\end{eqnarray}
Here $b$ is the Nakanishi-Lautrup auxiliary variable which linearizes the gauge fixing term $\displaystyle-\frac{1}{2}\left[\dot{\lambda} + P_\theta + P_\phi \right]^{2}$ and $(\bar{c})c$ are anti-commuting (anti-)ghost variables.  Now, expressing the above Lagrangian \eqref{L_brst} in the explicit form
\begin{eqnarray} \label{L_brst_1}
 L_b &=& \dot{\eta}P_\eta + \dot{\theta}P_\theta + \dot{\phi}P_\phi  + \lambda(p\theta + q\phi)  - b (\dot{\lambda} + P_\theta + P_\phi) + \frac{1}{2}b^{2}  \\ \nonumber 
&-& \frac{(\cosh\eta - \cos\theta)^2}{2ma^2}\Big(P_\eta^2 + \frac{\left( qP_{\theta}- pP_{\phi}\right)^{2}}{\left( p^{2}\sinh^{2}\eta + q^{2}\right)}\Big) + i \dot{\bar{c}}\dot{c} - i (p + q) \bar c c.
\end{eqnarray}
The above first-order Lagrangian \eqref{L_brst_1} is also invariant under the following anti-BRST symmetry transformations $(s_{ab})$ 
\begin{eqnarray} \label{brst}
	&& s_{ab}\lambda \;=\; \dot{\bar c}, \quad s_{ab} P_{\theta} \;=\; {\bar c}p, \quad s_{ab} P_{\phi} \;=\; {\bar c}q, \quad	s_{ab}c \;=\; -ib, \\ \nonumber
	&& s_{ab} \bar{c} \;=\; 0, \quad s_{ab} b \;=\; s_{ab} \eta \;=\;  s_{ab} P_{\eta} \;=\;  s_{ab} \theta \;=\;  s_{ab} \phi \;=\;  0.
\end{eqnarray}
The two key features of the (anti-)BRST symmetry transformations are: the off-shell nilpotency (i.e. $s_{b}^{2} = 0 = s_{ab}^{2} $) and absolute anti-commutativity (i.e. $\left\lbrace s_{b}, s_{ab} \right\rbrace  =  s_{b}s_{ab} + s_{ab}s_{b} = 0$). 
The conserved (anti-)BRST charges ($Q_{(a)b}$) are deduced as
\begin{equation}
	Q_{b} \;=\; -c(p\theta+q\phi) + \dot{c}P_{\lambda}, \quad Q_{ab} \;=\; -\bar{c}(p\theta+q\phi) + \dot{\bar{c}}P_{\lambda}.
\end{equation}
These (anti-)BRST charges are nilpotent of order two (i.e.  $Q_{b}^{2} = 0 = Q_{ab}^{2}$) and absolutely anti-commuting ($\lbrace Q_{b}, Q_{ab}\rbrace = Q_{b}Q_{ab}+Q_{ab}Q_{b}=0$) in nature.  Moreover, these charges act as the generators of (anti-)BRST symmetry transformations which can be verified with the aid of $s_{(a)b}\psi = i\left[ Q_{(a)b}, \psi \right] _{\pm},$ where  $\psi$ is any generic variable and $\pm$ represents the  (anti-)commutation relation.

\section{Supervariable approach: (Anti-)BRST symmetries}
In this section, we explore the model of particle confined to move on a $(p, q)$-torus knot within the framework of supervariable approach. We capture the full set of (anti-)BRST symmetry transformations and provide a geometrical explication regarding their properties.  In the supervariable approach,  we start with specifying the dynamics in terms of the coordinates\footnote{Ultimately, we shall take the limit 
$\eta \rightarrow 0$ and $\phi \rightarrow 0$, so that all the variables become a function of evolution parameter $t$ only.} $\eta$, $\phi$ and evolution parameter $t$ as the system is subjected to constrained condition $p \theta + q \phi \approx 0$. Thus, the exterior derivative is defined as follows
\begin{eqnarray}
    d &=& dt\partial_{t} + d\eta\partial_{\eta} + d\phi\partial_{\phi},
\end{eqnarray}
with the following properties:
\begin{eqnarray}
    dt \wedge d\eta \;=\; -  d\eta \wedge  dt, \quad dt \wedge d\phi \;=\; -  d\phi \wedge  dt, \quad d\eta \wedge d\phi \;=\; -  d\phi \wedge  d\eta.
\end{eqnarray}
The components of gauge potentials in the theory are $\lambda$, $P_{\theta}$ and $P_{\phi}$ , by virtue of the fact that the first-class constraints are $P_{\lambda} \approx 0$ and $(p\theta + q\phi) \approx 0$. The canonical one-form is then defined as
\begin{eqnarray}
   A^{(1)} &=& dt\lambda(\eta, \phi, t) + d\eta B(\eta, \phi, t) + d\phi D(\eta, \phi, t),
\end{eqnarray}
where the components $B$ and $D$ would be connected to the gauge potentials $P_{\theta}$ and $P_{\phi}$ respectively, through appropriate choices. 
Now the curvature two-form is constructed by taking the exterior derivative of one-form $A^{(1)}$ as
\begin{eqnarray}
    d  A^{(1)} &=& (dt \wedge d\eta )(\partial_{t}B - \partial_{\eta}\lambda) + (dt \wedge d\phi) (\partial_{t}D - \partial_{\phi}\lambda) + (d\eta \wedge d\phi) (\partial_{\eta}D - \partial_{\phi}B).
\end{eqnarray}
This curvature two-form is invariant under gauge and (anti-)BRST symmetry transformations. Now, in supervariable approach, we enlarge the original manifold of commuting variables $(\eta, \phi, t)$ by introducing a pair of anti-commuting Grassmannian variables $\beta$ and $\bar{\beta}$ (where $\beta^{2} = 0$, $\bar{\beta}^{2} = 0$, $\beta\bar{\beta}+\bar{\beta}\beta = 0$). The generalized forms of  the super exterior derivative ($\tilde{d}$) and super one-form ($\tilde{A}^{(1)}$) onto this $(3,2)$-dimensional supermanifold,  respectively, are given below 
\begin{eqnarray}
   d \;\rightarrow \;\tilde{d} &=& dt\partial_{t} + d\eta\partial_{\eta} + d\phi\partial_{\phi} + d\beta\partial_{\beta} + d\bar{\beta}\partial_{\bar{\beta}}, \\ \nonumber
   A^{(1)} \; \rightarrow \; \tilde{A}^{(1)} &=& dt\tilde{\lambda}(\eta, \phi, t, \beta, \bar{\beta}) + d\eta \tilde{B}(\eta, \phi, t, \beta, \bar{\beta}) + d\phi \tilde{D}(\eta, \phi, t, \beta, \bar{\beta}) \\ \nonumber
   &+&  d\beta \bar{F}(\eta, \phi, t, \beta, \bar{\beta}) +  d\bar{\beta} F(\eta, \phi, t, \beta, \bar{\beta}).
\end{eqnarray}
Here, the $(\bar{F})F$ are supervariable analogues of the ordinary (anti-)ghost variables ($(\bar{c})c$). We now expand the supervariables along the Grassmannian direction as
\begin{eqnarray} \label{super_exp}
 \tilde{\lambda} (\eta, \phi, t, \beta, \bar{\beta}) &=& \lambda (\eta, \phi, t) +\beta \bar{f}_{1}(\eta, \phi, t)  +\bar{\beta} f_{1}(\eta, \phi, t) +  i\beta\bar{\beta} B_{1}(\eta, \phi, t),  \\ \nonumber
 \tilde{B} ( \eta, \phi, t, \beta, \bar{\beta}) &=& B(\eta, \phi, t) +\beta \bar{f}_{2}(\eta, \phi, t)  +\bar{\beta} f_{2}(\eta, \phi, t)  + i\beta\bar{\beta} B_{2}(\eta, \phi, t), \\ \nonumber
 \tilde{D}(\eta, \phi, t, \beta, \bar{\beta}) &=& D(\eta, \phi, t) +\beta \bar{f}_{3}(\eta, \phi, t)  +\bar{\beta} f_{3}(\eta, \phi, t) + i\beta\bar{\beta} B_{3}(\eta, \phi, t),  \\ \nonumber
 F (\eta, \phi, t, \beta, \bar{\beta}) &=& c (\eta, \phi, t) +i\beta \bar{b}_{1}(\eta, \phi, t)  +i\bar{\beta} b_{1}(\eta, \phi, t) + i\beta\bar{\beta} S(\eta, \phi, t),  \\ \nonumber
 \bar{F} (\eta, \phi, t, \beta, \bar{\beta}) &=& \bar{c} (\eta, \phi, t) +i\beta \bar{b}_{2}(\eta, \phi, t)  +i\bar{\beta} b_{2}(\eta, \phi, t)
+ i\beta\bar{\beta} \bar{S}(\eta, \phi, t),
\end{eqnarray}
where secondary variables $ f_{1}, \; \bar{f}_{1},\;  f_{2},\;  \bar{f}_{2}, \; f_{3}, \;\bar{f}_{3},\;  S, \; \bar{S}$ are fermionic in nature and $B_{1}, \;B_{2},\; B_{3},\; b_{1}, \; \bar{b}_{1},\;  b_{2},\; \bar{b}_{2}$ are bosonic. In the above expression, all the secondary variables would be expressed in terms of basic variables via implementing the horizontality condition on the curvature two-form. The horizontality condition demands that the curvature two-form, in the supermanifold,  remains unaffected by the presence of Grassmannian variables. In other words, we impose the following condition 
\begin{eqnarray}
     d  A^{(1)} &=& \tilde{d}  \tilde{A}^{(1)},
\end{eqnarray}
where the curvature two-form, in $(3,2)$-dimensional supermanifold, is given below
\begin{eqnarray}
    \tilde{d}  \tilde{A}^{(1)} &=& (dt \wedge d\eta) (\partial_{t}\tilde{B} - \partial_{\eta}\tilde{\lambda}) + (dt \wedge d\phi ) (\partial_{t}\tilde{D} - \partial_{\phi}\tilde{\lambda}) +  (d\eta \wedge d\phi ) (\partial_{\eta}\tilde{D} - \partial_{\phi}\tilde{B}) \\ \nonumber
    &+&   (dt \wedge d\beta) (\partial_{t}\bar{F} - \partial_{\beta}\tilde{\lambda}) + (dt \wedge d\bar{\beta}) (\partial_{t}F - \partial_{\bar{\beta}}\tilde{\lambda}) + (d\eta \wedge d\beta) (\partial_{\eta}\bar{F} - \partial_{\beta}\tilde{B}) \\ \nonumber
    &+& (d\eta \wedge d\bar{\beta}) (\partial_{\eta}F - \partial_{\bar{\beta}}\tilde{B}) + (d\phi \wedge d\beta ) (\partial_{\phi}\bar{F} - \partial_{\beta}\tilde{D}) + (d\phi \wedge d\bar{\beta}) (\partial_{\phi}F - \partial_{\bar{\beta}}\tilde{D})\\ \nonumber
    &-& (d\beta \wedge d\bar{\beta}) (\partial_{\beta}F + \partial_{\bar{\beta}}\bar{F}) - (d\beta \wedge d\beta )\partial_{\beta}\bar{F} - (d\bar{\beta} \wedge d\bar{\beta}) \partial_{\bar{\beta}}F . 
\end{eqnarray}
Now, by making use of the horizontality condition, we yield the following relations for the secondary variables in terms of the basic variables as
\begin{eqnarray}
   && f_{1} \;=\; \dot{c}, \quad \bar{f_{1}} \;=\; \dot{\bar{c}}, \quad f_{2} \;=\; \partial_{\eta} c, \quad \bar{f_{2}} \;=\; \partial_{\eta} \bar{c}, \\ \nonumber
 &&  f_{3} \;=\; \partial_{\phi} c, \quad \bar{f}_{3} \;=\; \partial_{\phi} \bar{c}, 
   \quad b_{1} \;=\; 0, \quad \bar{b}_{2} \;=\; 0, \\ \nonumber
  && S \;=\; 0, \quad  \bar{S} \;=\; 0, \quad B_{1} \;=\; \dot{b_{2}} \;=\; - \dot{\bar{b}}_{1}, \quad \bar{b}_{1}+b_{2} \;=\; 0,\\ \nonumber
  && B_{2} \;=\; \partial_{\eta}b_{2} \;=\; - \partial_{\eta}\bar{b}_{1}, \quad  B_{3} \;=\; \partial_{\phi}b_{2} \;=\; - \partial_{\phi}\bar{b}_{1}.
\end{eqnarray}
In the above, we choose $b_{2} = b = -\bar{b}_{1}$ from the relation $\bar{b}_{1}+b_{2} =0$, and substituting for these secondary variables in the expansion of supervariables \eqref{super_exp} produces the following explicit relations:
\begin{eqnarray} \label{sup_first}
     \tilde{\lambda}^{(R)} (\eta, \phi, t, \beta, \bar{\beta}) &=& \lambda (\eta, \phi, t) +\beta \dot{\bar{c}}(\eta, \phi, t)  +\bar{\beta} \dot{c}(\eta, \phi, t)  + i\beta\bar{\beta} \dot{b}(\eta, \phi, t),  \\ \nonumber
    \tilde{B}^{(R)} ( \eta, \phi, t, \beta, \bar{\beta}) &=& B(\eta, \phi, t) +\beta \partial_{\eta}\bar{c}(\eta, \phi, t)  +\bar{\beta} \partial_{\eta}c(\eta, \phi, t) + i\beta\bar{\beta} \partial_{\eta}b(\eta, \phi, t) , \\ \nonumber
     \tilde{D}^{(R)}(\eta, \phi, t, \beta, \bar{\beta}) &=& D(\eta, \phi, t) +\beta \partial_{\phi}\bar{c}(\eta, \phi, t)  +\bar{\beta} \partial_{\phi}c(\eta, \phi, t) +  i\beta\bar{\beta} \partial_{\phi}b(\eta, \phi, t), \\ \nonumber
      F^{(R)} (\eta, \phi, t, \beta, \bar{\beta}) &=& c (\eta, \phi, t) - i\beta b(\eta, \phi, t), \\ \nonumber
     \bar{F}^{(R)} (\eta, \phi, t, \beta, \bar{\beta}) &=& \bar{c} (\eta, \phi, t)   +i\bar{\beta} b(\eta, \phi, t).
\end{eqnarray} 
Here the superscript $(R)$, in the expansions, denotes the reduced supervariable forms. Now, we choose appropriate relations connecting the variables,  $B$ and $D$, with the gauge potentials $P_{\theta}( \eta, \phi, t)$ and $P_{\phi}( \eta, \phi, t)$, respectively
\begin{eqnarray}
    B ( \eta, \phi, t) \;=\; \frac{1}{p} \partial_{\eta}  P_{\theta} ( \eta, \phi, t), \quad \tilde{B}^{(R)} ( \eta, \phi, t, \beta, \bar{\beta}) \;=\; \frac{1}{p} \partial_{\eta}  \tilde{P}_{\theta}^{(R)} ( \eta, \phi, t, \beta, \bar{\beta}),\\ \nonumber
      D ( \eta, \phi, t) \;=\; \frac{1}{q} \partial_{\phi}  P_{\phi} ( \eta, \phi, t), \quad  \tilde{D}^{(R)} ( \eta, \phi, t, \beta, \bar{\beta}) \;=\; \frac{1}{q} \partial_{\phi}  \tilde{P}_{\phi}^{(R)} ( \eta, \phi, t, \beta, \bar{\beta}).
    \end{eqnarray} 
Substitute these relations into \eqref{sup_first}, we have
   \begin{eqnarray} 
     \tilde{P}_{\theta}^{(R)} ( \eta, \phi, t, \beta, \bar{\beta}) &=& P_{\eta}(\eta, \phi, t) +\beta p\bar{c}(\eta, \phi, t)  +\bar{\beta} pc(\eta, \phi, t) +  i\beta\bar{\beta} pb(\eta, \phi, t), \\ \nonumber
     \tilde{P}_{\phi}^{(R)}(\eta, \phi, t, \beta, \bar{\beta}) &=&  P_{\phi}(\eta, \phi, t) +\beta q\bar{c}(\eta, \phi, t)  +\bar{\beta} qc(\eta, \phi, t) + i\beta\bar{\beta} qb(\eta, \phi, t).
\end{eqnarray}
Since all the variables in our theory are functions of evolution parameter $t$; thus we take the limits $\eta \rightarrow 0$ and $\phi \rightarrow 0$ and obtain the following expansions for supervariables
\begin{eqnarray} \label{fin_supervar}
    \tilde{\lambda}^{(h)} (t, \beta, \bar{\beta}) &=& \lambda (t) +\beta \dot{\bar{c}}(t)  +\bar{\beta} \dot{c}(t) + i\beta\bar{\beta} \dot{b}(t),  \\ \nonumber
     \tilde{P}_{\theta}^{(h)} (t, \beta, \bar{\beta}) &=& P_{\theta}(t) +\beta p\bar{c}(t)  +\bar{\beta} pc(t) + i\beta\bar{\beta} pb(t), \\ \nonumber
     \tilde{P}_{\phi}^{(h)}(t, \beta, \bar{\beta}) &=&  P_{\phi}(t) +\beta q\bar{c}(t)  +\bar{\beta} qc(t) + i\beta\bar{\beta} qb(t), \\  \nonumber
     F ^{(h)}(t, \beta, \bar{\beta}) &=& c (t) - i\beta b(t),  \\ \nonumber
     \bar{F}^{(h)} (t, \beta, \bar{\beta}) &=& \bar{c} (t)   +i\bar{\beta} b(t).
\end{eqnarray}
Here, the superscript $(h)$ represents the supervariable expansion upon the implementation of the horizontality condition.  In terms of (anti-)BRST symmetry transformations, the above expansions can be rewritten as
\begin{eqnarray} \label{super_var_brst}
    \tilde{\lambda}^{(h)} (t, \beta, \bar{\beta}) &=& \lambda (t) +\beta (s_{ab}\lambda)  +\bar{\beta} (s_{b}\lambda) + \beta\bar{\beta} (s_{b}s_{ab}\lambda),  \\ \nonumber
    \tilde{P}_{\theta} ^{(h)}( t, \beta, \bar{\beta}) &=& P_{\theta}(t) +\beta (s_{ab}P_{\theta})  +\bar{\beta} (s_{b}P_{\theta}) + \beta\bar{\beta} (s_{b}s_{ab}P_{\theta}), \\ \nonumber
     \tilde{P}_{\phi}^{(h)} (t, \beta, \bar{\beta}) &=& P_{\phi}(t) +\beta (s_{ab}P_{\phi})  +\bar{\beta} (s_{b}P_{\phi}) + \beta\bar{\beta} (s_{b}s_{ab}P_{\phi}), \\ \nonumber
     F ^{(h)}(t, \beta, \bar{\beta}) &=& c (t) +\beta (s_{ab}c)  +\bar{\beta} (s_{b}c) + \beta\bar{\beta} (s_{b}s_{ab}c), \\ \nonumber
     \bar{F}^{(h)} (t, \beta, \bar{\beta}) &=& \bar{c} (t) +\beta (s_{ab}\bar{c})  +\bar{\beta} (s_{b}\bar{c}) + \beta\bar{\beta} (s_{b}s_{ab}\bar{c}).
\end{eqnarray}
In the super expansion of the (anti-)ghost variables, we incorporated the vanishing infinitesimal (anti-)BRST symmetries: $s_{b}c =0$ and $s_{ab}\bar{c} = 0$. This provides the non-trivial (anti-)BRST symmetries.

To derive the trivial (anti-)BRST symmetries: i.e., $s_{(a)b}\theta = 0$, $s_{(a)b}\phi = 0$, $s_{(a)b}\eta = 0$ and $s_{(a)b}P_{\eta} = 0$, we utilizes the ``augmented" supervariable approach. In this approach, the supervariables corresponding to the gauge invariant (BRST invariant) dynamical variables are treated to be independent of Grassmannian variables on generalizing onto the $(1,2)$-dimensional supermanifold which is parameterized by $(t, \beta, \bar{\beta})$.  This condition can be expressed as follows:
\begin{eqnarray} \label{tri_brst}
    &&\tilde{\theta} (t, \beta, \bar{\beta}) \;=\; \theta(t), \quad \tilde{\phi} (t, \beta, \bar{\beta}) \;=\; \phi(t), \\ \nonumber 
   && \tilde{\eta} (t, \beta, \bar{\beta}) \;=\; \eta(t), \quad  \tilde{P_{\eta}} (t, \beta, \bar{\beta}) \;=\; P_{\eta}(t).
\end{eqnarray}
Subsequently, the expansion of $\theta, \phi, \eta$ and $ P_{\eta}$ onto the $(1,2)$-dimensional supermanifold produces the following expressions
\begin{eqnarray}
    \tilde{\theta} (t, \beta, \bar{\beta}) &=& \theta (t) +\beta \bar{f}_{4}(t)  +\bar{\beta} f_{4}(t) + i\beta\bar{\beta} B_{4}(t),  \\ \nonumber
    \tilde{\phi} (t, \beta, \bar{\beta}) &=& \phi (t) +\beta \bar{f}_{5}(t)  +\bar{\beta} f_{5}(t) + i\beta\bar{\beta} B_{5}(t),  \\ \nonumber
    \tilde{\eta} (t, \beta, \bar{\beta}) &=& \eta (t) +\beta \bar{f}_{6}(t)  +\bar{\beta} f_{6}(t) + i\beta\bar{\beta} B_{6}(t),  \\ \nonumber
    \tilde{P_{\eta}} (t, \beta, \bar{\beta}) &=& P_{\eta} (t) +\beta \bar{f}_{7}(t)  +\bar{\beta} f_{7}(t) + i\beta\bar{\beta} B_{7}(t),
\end{eqnarray}
where $f_{4}, \bar{f}_{4},f_{5}, \bar{f}_{5},f_{6}, \bar{f}_{6},f_{7}, \bar{f}_{7}$ are fermionic secondary variables and $B_{4}, B_{5}, B_{6}, B_{7}$ represent the bosonic secondary variables. The condition \eqref{tri_brst} furnishes the following expressions for the secondary variables
\begin{eqnarray}
    &f_{4} \;=\; \bar{f}_{4} \;=\; f_{5} \;=\; \bar{f}_{5} \;=\; f_{6} \;=\; \bar{f}_{6} \;=\; f_{7} \;=\; \bar{f}_{7} \;=\; 0, &\\ \nonumber
   & B_{4} \;=\;  B_{5} \;=\;  B_{6} \;=\;  B_{7} \;=\; 0.&
\end{eqnarray}
Consequently, the expansion of variables (i.e. $\theta,  \phi,  \eta, P_{\eta}$), onto the supermanifold is obtained as
\begin{eqnarray}
     \tilde{\theta} (t, \beta, \bar{\beta}) &=& \theta (t) +\beta (s_{ab}\theta) +\bar{\beta} (s_{b}\theta) + \beta\bar{\beta} (s_{b}s_{ab}\theta),  \\ \nonumber
     \tilde{\phi} (t, \beta, \bar{\beta}) &=& \phi (t) +\beta (s_{ab}\phi) +\bar{\beta} (s_{b}\phi) + \beta\bar{\beta} (s_{b}s_{ab}\phi),  \\ \nonumber
     \tilde{\eta} (t, \beta, \bar{\beta}) &=& \eta (t) +\beta (s_{ab}\eta) +\bar{\beta} (s_{b}\eta) + \beta\bar{\beta} (s_{b}s_{ab}\eta),  \\ \nonumber
     \tilde{P}_{\eta} (t, \beta, \bar{\beta}) &=& P_{\eta} (t) +\beta (s_{ab}P_{\eta}) +\bar{\beta} (s_{b}P_{\eta}) + \beta\bar{\beta} (s_{b}s_{ab}P_{\eta})  ,
\end{eqnarray}
where we have employed the trivial (anti-)BRST symmetry transformations to obtain the above expressions. 

\section{Key features of (anti-)BRST symmetries: Supervariable approach}
In this section, we capture some of the key features of (anti-)BRST symmetries within the supervariable approach.  In this formalism,  the translational along the Grassmannian direction $\bar{\beta}$ (i.e., $\frac{\partial}{\partial \bar{\beta}}$), for any generic supervariable $\bar{\Psi}(t, \beta, \bar{\beta})$ in the limit as $\beta$ to zero,  results the  BRST symmetry of the corresponding dynamic variable. Similarly, the translation operator $\frac{\partial}{\partial \beta}$ (along the Grassmannian direction $\beta$) operating on the generic supervariable, by keeping other Grassmannian parameter ($\bar{\beta}$) tending to zero, produces the anti-BRST symmetry. On the other hand, the translation along the Grassmannian direction $\beta$(or $\bar{\beta}$) followed by a translational along $\bar{\beta}$(or $\beta$) on any supervariable generate $s_{b}s_{ab}\Psi(t)$[or $s_{ab}s_{b}\Psi(t)$]. These relationships are expressed in the following manner
\begin{eqnarray}
    &\displaystyle \Lim{\beta \rightarrow 0} \frac{\partial}{\partial \bar{\beta}} \bar{\Psi}(t, \beta, \bar{\beta}) \;=\; s_{b}\Psi(t), \quad \displaystyle \Lim{\bar{\beta} \rightarrow 0} \frac{\partial}{\partial \beta} \bar{\Psi}(t, \beta, \bar{\beta}) \;=\; s_{ab}\Psi(t),& \\ \nonumber 
 &\displaystyle \frac{\partial}{\partial \bar{\beta}}\frac{\partial}{\partial \beta}  \bar{\Psi}(t, \beta, \bar{\beta}) \;=\; s_{b}s_{ab}\Psi(t),&
\end{eqnarray}
where $\Psi(t)$ encompass of [$\eta, \theta, \phi, P_{\eta}, P_{\theta}, P_{\phi}, \lambda, c, \bar{c}$]. As it is evident that the (anti-)BRST symmetries satisfy two key characteristics such as: nilpotency of order two and absolute anti-commutativity.  Thus, the two successive operations of translational derivatives on any Grassmannian direction, as $\frac{\partial}{\partial \bar{\beta}}$ or $\frac{\partial}{\partial \beta}$  acting on the supervariable, furnishes the nilpotency of the BRST and anti-BRST symmetries, respectively,  as mathematically illustrated below 
\begin{eqnarray}
    &&\Lim{\beta \rightarrow 0} \frac{\partial}{\partial \bar{\beta}}\frac{\partial}{\partial \bar{\beta}}  \bar{\Psi}(t, \beta, \bar{\beta})\;=\;0 \;\Longleftrightarrow\; s_{b}^{2}\Psi(t)\;=\;0, \\ \nonumber
&& \Lim{\bar{\beta} \rightarrow 0} \frac{\partial}{\partial \beta}\frac{\partial}{\partial \beta} \bar{\Psi}(t, \beta, \bar{\beta})\;=\;0 \;\Longleftrightarrow\; s_{ab}^{2}\Psi(t)\;=\;0.
\end{eqnarray}
Similarly, the {\it{sum}} of action of the translational generator along $\beta$ followed by the same along $\bar{\beta}$ and translational generator along $\bar{\beta}$ followed by the one along $\beta$ establishes the absolute anti-commutativity of $s_{b}$ and $s_{ab}$ (i.e. $\{s_{b}s_{ab}+s_{ab}s_{b}\} = 0$) in the following manner
\begin{eqnarray}
 \left(\frac{\partial}{\partial \bar{\beta}}\frac{\partial}{\partial \beta} + \frac{\partial}{\partial \beta}\frac{\partial}{\partial \bar{\beta}} \right) \bar{\Psi}(t, \beta, \bar{\beta})\;=\;0 \;\Longleftrightarrow\; \left(s_{b}s_{ab}+s_{ab}s_{b}\right)\Psi(t) \;=\;0.
\end{eqnarray}
Moreover, we can also capture the nilpotency and absolute anti-commutativity of the (anti-)BRST charges with the aid of the Grassmannian translational operators. In the process of providing an analog in the supervariable approach, we express the BRST and anti-BRST charges onto the $(1, 2)$-dimensional supermanifold. To begin with,  the BRST charge $Q_{b}$ can be written in the following forms
\begin{eqnarray}
 Q_{b} \;=\; is_{b}\left(c\dot{\bar{c}}- \dot{c}\bar{c}\right) \;=\; is_{ab}\left(\dot{c}c\right) \;=\; -is_{b}s_{ab}\left(\lambda c\right).
\end{eqnarray}
We have used the fact that $P_{\lambda} = -b$ and $\dot{b} = -(p\theta + q\phi)$ arising from the definition of momentum and equations of motion from Lagrangian \eqref{L_brst_1} (cf. \cite{gupt} for details). Now, making use of the supervariable expansions (cf.  \eqref{fin_supervar}) and analog of the (anti-)BRST symmetries in terms of the Grassmannian operators, we generalize the above expressions for the BRST charge onto the $(1, 2)$-dimensional supermanifold in the following manner
\begin{eqnarray}
 Q_{b} &=& i\frac{\partial}{\partial \bar{\beta}}\left(F^{(h)}\dot{\bar{F}}^{(h)} - \dot{F}^{(h)}\bar{F}^{(h)}\right)\Big|_{\beta = 0} \;\equiv\; i\int d\bar{\beta}\left(F^{(h)}\dot{\bar{F}}^{(h)} - \dot{F}^{(h)}\bar{F}^{(h)}\right)\Big|_{\beta = 0}, \\ \nonumber
&=& i\frac{\partial}{\partial \beta}\left(\dot{F}^{(h)}F^{(h)}\right) \;\equiv\; i\int d \beta\left(\dot{F}^{(h)}F^{(h)}\right), \\ \nonumber
&=& -i\frac{\partial}{\partial \bar{\beta}}\frac{\partial}{\partial \beta}\left(\tilde{\lambda}^{(h)}F^{(h)}\right) \;\equiv\; -i\int d\bar{\beta}d \beta \left(\tilde{\lambda}^{(h)}F^{(h)}\right).
\end{eqnarray}
In the similar manner, we express the anti-BRST charge in terms of (anti-)BRST symmetries as
\begin{eqnarray}
 Q_{ab} \;=\; -is_{ab}\left(c\dot{\bar{c}}- \dot{c}\bar{c}\right) \;=\; -is_{b}\left(\dot{\bar{c}}\bar{c}\right) \;=\; -is_{b}s_{ab}\left(\lambda \bar{c}\right).
\end{eqnarray}
Here also we have utilized expression for momentum, $P_{\lambda} = -b$ and the equation of motion, $\dot{b} = -(p\theta + q\phi)$ from the BRST invariant Lagrangian \eqref{L_brst_1} (cf. \cite{gupt} for details). Likewise, as for the BRST charge, we generalize the above expressions for the anti-BRST charge onto the $(1, 2)$-dimensional supermanifold in the following fashion
\begin{eqnarray}
 Q_{ab} &=& -i\frac{\partial}{\partial \beta}\left(F^{(h)}\dot{\bar{F}}^{(h)} - \dot{F}^{(h)}\bar{F}^{(h)}\right)\Big|_{\bar{\beta} = 0} \;\equiv\; -i\int d \beta\left(F^{(h)}\dot{\bar{F}}^{(h)} - \dot{F}^{(h)}\bar{F}^{(h)}\right)\Big|_{\bar{\beta} = 0}, \\ \nonumber
&=& -i\frac{\partial}{\partial \bar{\beta}}\left(\dot{\bar{F}}^{(h)}\bar{F}^{(h)}\right) \;\equiv\; -i\int d\bar{\beta}\left(\dot{\bar{F}}^{(h)}\bar{F}^{(h)}\right), \\ \nonumber
&=& -i\frac{\partial}{\partial \bar{\beta}}\frac{\partial}{\partial \beta}\left(\tilde{\lambda}^{(h)}\bar{F}^{(h)}\right) \;\equiv\; -i\int d\bar{\beta}d \beta \left(\tilde{\lambda}^{(h)}\bar{F}^{(h)}\right).
\end{eqnarray}
It is worthwhile to mention that,  from the above expressions, we can demonstrate in a straightforward manner that $s_{b}Q_{b} = 0$ and $s_{ab}Q_{ab} = 0$, which in turn implicate the nilpotency nature of (anti-)BRST charges (i.e.,  $Q_{b}^{2}=0$ and $Q_{ab}^{2}=0$). Moreover, the relations - $s_{ab}Q_{b} = 0$ and $s_{b}Q_{ab} = 0$, establish the absolute anti-commutativity ($Q_{b}Q_{ab}+Q_{ab}Q_{b}=0$). Thus, we can provide an interpretation of these features with the aid of translational operator along Grassmannian variables as
\begin{eqnarray}
    &\frac{\partial}{\partial \bar{\beta}} Q_{b}\;=\; 0 \;\Longleftrightarrow\;Q_{b}^{2} \;=\; 0, \quad \frac{\partial}{\partial \beta} Q_{ab} \;=\; 0 \;\Longleftrightarrow\; Q_{ab}^{2} \;=\; 0 ,& \\ \nonumber 
 &\frac{\partial}{\partial \bar{\beta}} Q_{ab}\;=\; \frac{\partial}{\partial \beta} Q_{b}\;=\; 0  \;\Longleftrightarrow\; Q_{b}Q_{ab}+ Q_{ab}Q_{b} \;=\; 0.&
\end{eqnarray}
Before ending this section,  we can present an elucidation on the invariance of the first-order (anti-)BRST invariant Lagrangian by employing the Grassmannian operators ($\frac{\partial}{\partial \beta}$, $\frac{\partial}{\partial \bar{\beta}}$). For this purpose, we generalize the first-order Lagrangian onto the $(1,  2)$-dimensional supermanifold by making use of non-trivial supervariable expansions as
\begin{eqnarray} \label{L_brst_super}
\tilde{L}_b &=& \dot{\eta}P_\eta + \dot{\theta}\tilde{P}_\theta^{(h)} + \dot{\phi}\tilde{P}_\phi^{(h)}  + \tilde{\lambda}^{(h)}(p\theta + q\phi)  - b (\dot{\tilde{\lambda}}^{(h)} + \tilde{P}_\theta^{(h)}  + \tilde{P}_\phi^{(h)} ) + \frac{1}{2}b^{2}  \\ \nonumber 
&-& \frac{(\cosh\eta - \cos\theta)^2}{2ma^2}\Big(P_\eta^2 + \frac{\left( q\tilde{P}_{\theta}^{(h)} - p\tilde{P}_{\phi}^{(h)} \right)^{2}}{\left( p^{2}\sinh^{2}\eta + q^{2}\right)}\Big) + i \dot{\bar{F}}^{(h)} \dot{F}^{(h)}  - i (p + q) \bar{F}^{(h)}  F^{(h)} .
\end{eqnarray}
Now,  in terms of the translational generators along $\bar{\beta}$ and $\beta$, we obtain the following relationships
\begin{eqnarray}
    \frac{\partial}{\partial \bar{\beta}} \tilde{L}_b \Big|_{\beta = 0}\;=\; \frac{d}{dt}\left[c\left(p\theta+q\phi\right) - b\dot{c}\right] \;\Longleftrightarrow\; s_{b}L_{b} \;=\; \frac{d}{dt}\left[c\left(p\theta+q\phi\right) - b\dot{c}\right], \ \\ \nonumber 
  \frac{\partial}{\partial \beta} \tilde{L}_b \Big|_{\bar{\beta} = 0}\;=\; \frac{d}{dt}\left[\bar{c}\left(p\theta+q\phi\right) - b\dot{\bar{c}}\right] \;\Longleftrightarrow\; s_{ab}L_{b} \;=\; \frac{d}{dt}\left[\bar{c}\left(p\theta+q\phi\right) - b\dot{\bar{c}}\right].
\end{eqnarray}
Here the operation of translational derivative along the ($\beta$)$\bar{\beta}$ onto the superspace Lagrangian ($\tilde{L}_b$) in the limit of ($\bar{\beta}$)$\beta$ goes to zero, generates the quasi-invariance of the first-order Lagrangian \eqref{L_brst_1}  under (anti-)BRST symmetries.

\section{(Anti-)co-BRST symmetries}
In this section we show that, in addition to the infinitesimal (anti-)BRST symmetries, the system of particle on a torus knot possesses additional fermionic symmetries, the (anti-)co-BRST symmetries [$s_{(a)d}$]. These off-shell nilpotent and absolutely anti-commuting (anti-)co-BRST symmetry transformations are given below:
\begin{eqnarray}
     & s_{d}\lambda \;=\; (p+q)\bar{c}, \quad s_{d} P_{\theta} \;=\; -p\dot{\bar{c}}, \quad s_{d} P_{\phi} \;=\; -q\dot{\bar{c}}, \quad	 s_{d} \bar{c} \;=\; 0, &\\ \nonumber
    & s_{d} c \;=\; i(p\theta+q\phi), \quad s_{d} \eta \;=\; s_{d} P_{\eta} \;=\;  s_{d} \theta \;=\; s_{d} \phi  \;=\; s_{d} b  \;=\; 0, & \\ \nonumber 
& s_{ad}\lambda \;=\; (p+q)c, \quad s_{ad} P_{\theta} \;=\; -p\dot{c}, \quad s_{ad} P_{\phi} \;=\; -q\dot{c}, \quad s_{ad}c \;=\; 0,&\\  \nonumber
& s_{ad} \bar{c} \;=\; -i(p\theta+q\phi), \quad s_{ad}\eta \;=\; s_{ad}P_{\eta} \;=\; s_{ad}\theta \;=\;  s_{ad}\phi  \;=\; s_{ad}b \;=\; 0.&
\end{eqnarray}
Here the gauge fixing term is invariant under $s_{(a)d}$, i.e. $s_{(a)d}\left[\dot{\lambda} + P_{\theta} + P_{\phi}\right] =0$. Moreover, the first-order Lagrangian \eqref{L_brst_1} also remains invariant under the (anti-)co-BRST symmetries as 
\begin{equation}
    s_{d} L_{b} \;=\; 0, \quad s_{ad} L_{b} \;=\; 0 .
\end{equation}
Now, according to the Noether's theorem, the presence of infinitesimal continuous symmetry transformations that leave the first-order Lagrangian invariant lead to the following conserved (anti-)co-BRST charges ($Q_{(a)d}$) as:
\begin{equation}
    Q_{d} \;=\; (p+q)P_{\lambda}\bar{c} + (p\theta+q\phi)\dot{\bar{c}}, \quad Q_{ad} \;=\; (p+q)P_{\lambda}c + (p\theta+q\phi)\dot{c}.
\end{equation}
These conserved charges $Q_{(a)d}$  act as the generators of the (anti-)co-BRST symmetries and it can be checked by making use of the following relation
\begin{eqnarray}
    s_{(a)d} \Psi &=& i \left[Q_{(a)d},  \Psi  \right]_{\pm}.
\end{eqnarray}
In the above,  $\Psi$ represents any generic variable of the theory and subscript ($\pm$) on the square bracket denotes (anti-)commutator depending on the (fermionic)bosonic nature of the variable. In the explicit forms, the non-trivial (anti-)commutators of co-BRST charge with the generic variables are
\begin{eqnarray}
     && s_{d} \lambda \;=\; i \left[Q_{d},  \lambda  \right]_{-} \;=\; (p+q)\bar{c}, \quad  s_{d} P_{\theta} \;=\; i \left[Q_{d},  P_{\theta}  \right]_{-} \;=\; -p\dot{\bar{c}}, \\ \nonumber
     && s_{d} P_{\phi} \;=\; i \left[Q_{d},  P_{\phi}  \right]_{-} \;=\; -q\dot{\bar{c}}, \quad s_{d} c \;=\; i \left[Q_{d},  c \right]_{+} \;=\; i(p\theta+q\phi).
\end{eqnarray}
This demonstrates that $Q_{d}$ acts as the generator of co-BRST symmetry $s_{d}$. On the other hand, the non-vanishing (anti-)commutators of the anti-co-BRST charge with generic variables take the following form 
\begin{eqnarray}
     && s_{ad} \lambda = i \left[Q_{ad},  \lambda  \right]_{-} = (p+q)c, \quad  s_{ad} P_{\theta} = i \left[Q_{ad},  P_{\theta}  \right]_{-} = -p\dot{c}, \\ \nonumber
     && s_{ad} P_{\phi} = i \left[Q_{ad},  P_{\phi}  \right]_{-} = -q\dot{c}, \quad s_{ad} \bar{c} = i \left[Q_{ad},  \bar{c} \right]_{+} = -i(p\theta+q\phi).
\end{eqnarray}
Correspondingly, the above expressions show that the $Q_{ad}$ acts as the generator of $s_{ad}$. However, on the other hand, the following anti-commutators among the (anti-)co-BRST charges
\begin{eqnarray}
    && s_{d} Q_{d} \;=\; i \{Q_{d}, Q_{d} \} \;=\; 0, \quad  s_{d} Q_{ad} \;=\; i \{Q_{d}, Q_{ad} \} \;=\; 0, \\ \nonumber
    && s_{ad} Q_{d} \;=\; i \{Q_{ad}, Q_{d} \} \;=\; 0, \quad  s_{ad} Q_{ad} \;=\; i \{Q_{ad}, Q_{ad} \} \;=\; 0,
\end{eqnarray}
indicate our conserved charges (also the corresponding (anti-)co-BRST symmetries) are nilpotent of order two (i.e., $Q_{(a)d}^{2} = 0$ or $s_{(a)d}^{2} = 0$) and absolutely anti-commuting (i.e., $\{Q_{d}, Q_{ad} \} \;=\; 0$ or $\{s_{d}, s_{ad} \} \;=\; 0$) in nature.

\section{Bosonic and ghost scale symmetries}
In the previous sections, we have established the four sets of fermionic symmetries, i.e.  (anti-)BRST and (anti-)co-BRST symmetries. Now,  the anti-commutator of BRST and co-BRST symmetries generates a bosonic symmetry ($s_{w}$), as
\begin{eqnarray}
    \{s_{b}, s_{d} \} &=& s_{w}.
\end{eqnarray}
The explicit forms of these transformations, acting on the variables of the theory, are as described below
\begin{eqnarray}
 & s_{w}\lambda \;=\; i [(p+q)b + p\dot{\theta}+ q\dot{\phi}], \quad s_{w}P_{\theta} \;=\; ip[-\dot{b} + p\theta+ q\phi], &\\ \nonumber
  & s_{w}P_{\phi} \;=\; iq[-\dot{b} + p\theta+ q\phi], \quad  s_{w} c \;=\; 0,\quad s_{w} \bar{c} \;=\; 0, \quad s_{w} b \;=\; 0,& \\ \nonumber
  & s_{w} \eta \;=\; 0,\quad s_{w} P_{\eta} \;=\; 0, \quad s_{w} \theta \;=\; 0,\quad s_{w} \phi  \;=\; 0.&
\end{eqnarray}
It is worthwhile to mention here that under this bosonic symmetry, the ghost sector of the Lagrangian remains invariant. Moreover, we realize that our first-order Lagrangian $L_{b}$ is quasi-invariant under $s_{w}$. 
 To be specific, under $s_{w}$ the Lagrangian $L_{b}$ transforms as
\begin{eqnarray}
    s_{w}L_{b} &=& \frac{d}{dt}\left[ -i \left(b(p\dot{\theta}+q\dot{\phi}) - (p\theta+q\phi)^{2} \right)\right],
\end{eqnarray}
which in turn demonstrates that the action integral is invariant. On the other hand, there is an additional non-vanishing anti-commutator between anti-BRST and anti-co-BRST symmetries as outlined below
\begin{eqnarray}
    \{s_{ab}, s_{ad} \} &=& s_{\bar{w}}.
\end{eqnarray}
In the definitive fashion, $s_{\bar{w}}$ acting on the dynamical variables are described below
\begin{eqnarray}
 & s_{\bar{w}}\lambda \;=\; -i [(p+q)b + p\dot{\theta}+ q\dot{\phi}], \quad s_{\bar{w}}P_{\theta} \;=\; -ip[-\dot{b} + p\theta+ q\phi],& \\ \nonumber
 & s_{\bar{w}}P_{\phi} \;=\; -iq[-\dot{b} + p\theta+ q\phi], \quad s_{\bar{w}} c \;=\; 0, \quad s_{\bar{w}} \bar{c} \;=\; 0, \quad s_{\bar{w}} b \;=\; 0, & \\ \nonumber
   & s_{\bar{w}} \eta \;=\; 0,\quad s_{\bar{w}} P_{\eta} \;=\; 0, \quad s_{\bar{w}} \theta \;=\; 0,\quad s_{\bar{w}} \phi  \;=\; 0. &
\end{eqnarray}
This bosonic symmetry $s_{\bar{w}}$ transforms the first-order Lagrangian $L_{b}$ to a total derivative as
\begin{eqnarray}
    s_{\bar{w}}L_{b} &=& \frac{d}{dt}\left[ i \left(b(p\dot{\theta}+q\dot{\phi}) - (p\theta+q\phi)^{2} \right)\right],
\end{eqnarray}
and renders the action integral invariant. Moreover, these bosonic symmetries are found to be dependent on each other and we identify the given below relation
\begin{equation}
 s_{w} + s_{\bar{w}} \;=\; 0 \quad \Longrightarrow  \quad \{s_{b}, s_{d} \} \;=\; s_{w} \;=\;  -\{s_{ab}, s_{ad} \}.
\end{equation}
Now we infer this continuous bosonic symmetry implicates the existence of a conserved bosonic charge $Q_{w}$, via Noether's theorem:
\begin{eqnarray}
     Q_{w} &=& -i  [ (p+q)P_{\lambda}^{2} + (p\theta+q\phi)^{2}].
\end{eqnarray}
This charge acts as the generator of bosonic symmetry $s_{w}$ and can be easily shown to be conserved with the aid of Euler-Lagrange equations of motion. Moreover, apart from the anti-commutator of $s_{(a)b}$ and $s_{(a)d}$ that generate $s_{w}$, we obtain remaining anti-commutators 
\begin{equation}
     \{s_{b}, s_{ab} \} \;=\; 0, \quad  \{s_{b}, s_{ad} \} \;=\; 0, \quad  \{s_{d}, s_{ad} \} \;=\; 0, \quad  \{s_{d}, s_{ab} \} \;=\; 0,
\end{equation}
to be absolutely zero.

We further infer that our first-order Lagrangian $L_{b}$ is invariant under the following ghost-scale global symmetry transformations
\begin{eqnarray}
   && c \rightarrow e^{\epsilon} c, \quad  \bar{c} \rightarrow e^{-\epsilon} \bar{c}, \quad  \eta \rightarrow \eta, \quad \theta \rightarrow \theta, \quad \phi \rightarrow \phi, \\ \nonumber
   && P_{\eta} \rightarrow  P_{\eta}, \quad  P_{\theta} \rightarrow  P_{\theta}, \quad  P_{\phi} \rightarrow  P_{\phi}, \quad \lambda \rightarrow \lambda, \quad b \rightarrow b,
\end{eqnarray}
with $\epsilon$ as the global scale parameter. The infinitesimal forms of above symmetry transformations are given below
\begin{eqnarray}
 && s_{g}c \;=\; c, \quad s_{g}\bar{c} \;=\; -\bar{c}, \quad  s_{g}\lambda \;=\; 0, \quad   s_{g}\eta \;=\; 0, \quad   s_{g}P_{\eta} \;=\; 0,  \\ \nonumber
&&  s_{g}\theta \;=\; 0, \quad  s_{g}P_{\theta} \;=\; 0, \quad  s_{g}\phi \;=\; 0, \quad  s_{g}P_{\phi} \;=\; 0, \quad  s_{g}b \;=\; 0. 
\end{eqnarray}
In addition, we deduce the ghost charge ($Q_{g}$) corresponding to this ghost symmetry transformation as: $ Q_{g} \;=\; i  \left(\dot{\bar{c}}c + \dot{c}\bar{c} \right)$ and it acts as the generator of $s_{g}$. Moreover,  proof for the conservation of the charge can be provided by the use of equations of motion.

\section{Algebraic aspects: Charges and cohomological operators}
In the previous sections, we have procured six sets of continuous and infinitesimal symmetry transformations for the gauge invariant theory of particle constrained to move on a $(p,q)$-torus knot. We realize that the (anti-)BRST ($Q_{(a)b}$), (anti-)co-BRST ($Q_{(a)d}$), bosonic ($Q_{w}$) and ghost charges ($Q_{g}$) (and the corresponding symmetries) satisfy the following algebra:
\begin{eqnarray}
   && Q_{(a)b}^{2} \;=\; 0, \quad  Q_{(a)d}^{2} \;=\; 0,\quad  \{Q_{b}, Q_{ab} \} \;=\; 0, \quad  \{Q_{d}, Q_{ad} \} \;=\; 0,  \\ \nonumber
     &&  \{Q_{d}, Q_{ab} \} \;=\; 0, \quad \{Q_{b}, Q_{ad} \} \;=\; 0, \quad i\left[Q_{g}, Q_{b}\right] \;=\; Q_{b}, \\ \nonumber
     &&i\left[Q_{g}, Q_{ab}\right] \;=\; -Q_{ab}, \quad i\left[Q_{g}, Q_{d}\right] \;=\; -Q_{d}, \quad i\left[Q_{g}, Q_{ad}\right] \;=\; Q_{ad}, \\ \nonumber
    && i\{Q_{b}, Q_{d} \} \;=\; Q_{w}, \quad -i\{Q_{ad}, Q_{ab} \} \;=\; Q_{w},  \quad [Q_{w}, Q_{r}] \;=\; 0,\;\; r\;=\; b,ab,d,ad . 
\end{eqnarray}
Here we made use of the (anti-)commutators, $[\eta, P_{\eta}] = i$, $[\theta, P_{\theta}] = i$, $[\phi, P_{\phi}] = i$, $[\lambda, P_{\lambda}] = i$,  $\{ c, \dot{\bar c} \}= 1$ and $\{ \bar{c}, \dot{c} \}= -1$ to obtain the above algebra. These algebraic structures of charges (symmetries) evoke a correlation with the algebra of de Rham cohomological operators of differential geometry. The algebra satisfied by the differential geometrical operators, the exterior derivative ($d$), co-exterior derivative ($\delta$) and Laplacian operator ($\Delta$) are given below  \cite{hod3}
\begin{eqnarray}
    &d^{2} \;=\; 0, \quad \delta^{2} \;=\; 0, \quad \Delta \;=\; \{d,\delta\} \;=\; d\delta+\delta d, \quad  [\Delta, d] \;=\; 0, \quad [\Delta, \delta] \;=\; 0. &
\end{eqnarray}
In a more elaborate fashion, the off-shell nilpotent (anti-)BRST ($Q_{(a)b}$) and  (anti-)co-BRST ($Q_{(a)d}$) charges can be mapped to nilpotent exterior ($d$) or co-exterior  ($\delta$) derivatives.  To be exact, we have the anti-commutator between the BRST and co-BRST charge as $ \{Q_{b}, Q_{d} \} = Q_{w}$ and analyzing it vis-\`{a}-vis the expression  $ \{d,\delta\} \;=\; \Delta$, we can draw the mapping for the $Q_{b}$, $Q_{d}$, $ Q_{w}$ to the exterior, co-exterior derivatives and Laplacian operator, respectively.  Moreover, as we have linked the $Q_{b}$ with the exterior derivative, now the anti-commutator $\{Q_{b}, Q_{ad} \} = 0$ allows us to also connect the $Q_{ad}$ to nilpotent exterior derivative ($d^{2}=0$). Similary,  taking into account of the fact that $Q_{d}$ is linked with co-exterior derivative ($\delta$) which is nilpotent of order two ($\delta^{2}=0$) enables us to map the $Q_{ab}$ with the $\delta$ since we have the relation $\{Q_{d}, Q_{ab} \} = 0$. With all these facts, we have the following maps among the conserved charges and cohomological operators
\begin{equation}
    (Q_{b},\; Q_{ad}) \Longrightarrow d, \quad   (Q_{d},\; Q_{ab}) \Longrightarrow \delta, \quad Q_{w} \Longrightarrow \Delta.
\end{equation}
The algebra satisfied by these (anti-)BRST, (anti-)co-BRST, bosonic and ghost symmetries and their corresponding conserved charges provides a physical realization of the de Rham cohomological operators of differential geometry. Therefore, the gauge invariant theory of particle moving on a $(p,q)$-torus knot portrays a toy model for Hodge theory.

\section{Conclusions}
In the present investigation, we have derived the off-shell nilpotent and absolutely anti-commuting (anti-)BRST symmetries within the framework of supervariable approach. We have also provided a geometrical interpretation for these symmetries with the aid of translational derivatives.  In order to accomplish our goal, we generalized the $3$-dimensional ordinary manifold onto a $(3,2)$-dimensional supermanifold and employed the horizontality condition. We have also captured the key characteristics of the (anti-)BRST symmetries (and their corresponding charges) within the framework of supervariable approach by making use of translational derivatives along the Grassmannian directions.

Further, we have demonstrated the existence of off-shell nilpotent and absolutely anti-commuting (anti-)co-BRST symmetries along with the corresponding conserved charges. These charges turn out as the generators of (anti-)co-BRST symmetries. Moreover, we have deduced the bosonic symmetry transformation, derived from the anti-commutator of these aforementioned (anti-)BRST and (anti-)co-BRST symmetries, which leaves the first-order Lagrangian quasi-invariant. In addition, we have determined the ghost symmetry transformations and procured the corresponding charge. Subsequently, we have established that the algebra satisfied by these conserved charges is analogous to the algebra of the de Rham cohomological operators of differential geometry. Thus, we have proved that the reformulated gauge invariant theory of particle constrained to move on a $(p, q)$-torus knot provides a prototype for Hodge theory.
.


\begin{thebibliography}{99}
\bibitem{gauge_Henn} M. Henneaux and C. Teitelboim, \textit{Quantization of Gauge Systems} (Princeton University Press, Princeton, 1992).
\bibitem{rotor} D. Nemeschansky, C. R. Preitschopf and M. Weinstein, \textit{Ann. Phys.} \textbf{183}, 226 (1988).
\bibitem{brst1} C. Becchi, A. Rouet and R. Stora, \textit{Phys. Lett.} B  \textbf{52}, 344 (1974).
\bibitem{brst2} C. Becchi, A. Rouet and R. Stora, \textit{Ann. Phys. (N. Y.)} \textbf{98}, 287 (1976).
\bibitem{brst3} I. V. Tyutin, \textit{Lebedev Institute Preprint,
Report No. FIAN-39 } (1975) (Unpublished).
\bibitem{super1} L. Bonora and M. Tonin, \textit{Phys. Lett. B} \textbf{98}, 48 (1981).
\bibitem{super2} L. Bonora, P. Pasti and M. Toni, \textit{Nouvo Cimento A} \textbf{63}, 353 (1981).
\bibitem{super3} R. Delbourgo and P. D. Jarvis,  \textit{J. Phys. A: Math. Gen.} \textbf{15},  611 (1982). 
\bibitem{sup3} S. Gupta and R. Kumar, \textit{Int. J. Mod. Phys.} A \textbf{31} 1650173 (2016).
\bibitem{sup4} R. Kumar and A. Shukla, \textit{Adv.  High Energy Phys.} \textbf{2018}, 7381387 (2018).
\bibitem{sup5} A. Pradeep, Anjali S, B. M. Nair and S. Gupta, \textit{Commun. Theor. Phys.} \textbf{72}, 105205 (2020). 
\bibitem{knot} C. C. Adams, \textit{The Knot Book: An Elementary Introduction to the Mathematical Theory of Knots} (American Mathematical Society, Providence, 2004).
\bibitem{wit} E. Witten, {\it Commun. Math. Phys.} {\bf 121}, 351 (1989).
\bibitem{fadd} L. Faddeev and A. J. Niemi, \textit{Nature} \textbf{387}, 58 (1997).
\bibitem{chern_insu} L. Gui-Geng, \textit{Nature} \textbf{609}, 925 (2022).
\bibitem{knt_rovelli} C. Rovelli and L. Smolin, \textit{Phys. Rev. Lett.} \textbf{61}, 1155 (1988).
\bibitem{int2} V. V. Sreedhar, \textit{Ann. Phys.} \textbf{359}, 20 (2015). 
\bibitem{ghosh} P. Das and S. Ghosh, \textit{Found. Phys.} \textbf{46}, 1649 (2016).
\bibitem{gupt} Anjali S and S. Gupta, \textit{Eur. Phys. J. Plus} \textbf{137}, 511 (2022).
\bibitem{neto} J. Ananias Neto, C. Neves and W. Oliveira, \textit{Phys. Rev. D} \textbf{63}, 085018 (2001).
\bibitem{nejad} S. A. Nejad, M. Dehghani and M. Monemzadeh, \textit{J. High Energy Phys.} \textbf{01}, 017 (2018).
\bibitem{hod4} R. P. Malik, \textit{Int. J. Mod. Phys.} A \textbf{22}, 3521 (2007). 
\bibitem{hod1} S. Gupta and R. P. Malik, \textit{Eur. Phys. J.} C \textbf{58}, 517 (2008).
\bibitem{hod2} S. Gupta and R. P. Malik, \textit{Eur. Phys. J.} C \textbf{68}, 325 (2010).
\bibitem{hod3} T. Eguchi, P. B. Gilkey and A. Hanson, \textit{Phys. Rep. } \textbf{66}, 213 (1980).
\end{thebibliography}
\end{document}